\documentstyle[aps,prl]{revtex}
\begin{document}
\title{Signatures of quantum chaos in nodal points and streamlines
in electron transport through billiards}
\author{Karl-Fredrik Berggren$^1$,
Konstantin N. Pichugin $^{2,3}$,
Almas F. Sadreev$^{1,2}$, and Anton Starikov$^2$}
\address{1) Department of Physics and Measurement Technology,
Link\"{o}ping University, S-581 83 Link\"{o}ping, Sweden\\
2) Kirensky Institute of Physics, 660036, Krasnoyarsk, Russia\\
3) Institute of Physics, Academy of Sciences, Cukrovarnick\'{a},
10, 16000 Prague}
\maketitle

\begin{abstract}

Streamlines and distributions of nodal points are used as signatures of chaos
in coherent electron transport through three types
of billiards, Sinai, Bunimovich and rectangular. Numerical averaged
 distribution  functions of nearest distances between nodal
points are presented. We find the same form for the Sinai and Bunimovich
 billiards and suggest that there is a universal form that can be used as
  a signature of quantum chaos for electron transport in open billiards.
The universal distribution function is found to be insensitive to the way
avaraging is performed (over positions of leads, over an energy interval
with a few conductance fluctuations, or both).
The integrable rectangular billiard, on the other hand, displays nonuniversal
 distribution
with a central peak related to partial order of nodal points for the
case of symmetric attachment of leads. However cases with nonsymmetric
leads tend to the universal form.

Also it is shown how nodal points in rectangular billiard can
lead to "channeling of quantum flows"
while disorder in nodal points in the Sinai billiard gives rise to
unstable irregular behavior of the flow.\noindent
\end{abstract}

\section{Introduction}
Billiards play a predominant role in the study of classical and quantum
chaos\cite{Guhr}.
Indeed, the nature of quantum chaos in a specific system is traditionally
inferred from its
its classical counterpart.
Hence one may ask if quantum chaos is to be understood solely as a phenomenon
that emerges in
the classical limit, or are there some intrinsically quantal phenomena,
which can contribute to irregular behavior in the quantum domain?
This is a question we raise in connection with
quantum transport through ideal regular and irregular electron billiards.

 The seminal studies by McDonald and Kauffmann \cite{McDonald} of the morphology
of eigenstates in a  closed Bunimovich stadium have
revealed characteristic patterns of disordered, undirectional
and non-crossing nodal
lines.
    Here we will first discuss what will happen to patterns like these
 when input and output leads are attached to a billiard, regular or irregular, and an electric
 current is induced through the
 the billiard by an applied voltage between the two leads. For such an open
 system
 the wave function $\psi$
 is now a scattering state with both real and imaginary parts, each of which gives rise
 to separate sets of nodal lines at which either $Re[\psi]$ or $Im[\psi]$
  vanish.  How will the patterns of nodal lines evolve as, e.g., the
  energy of injected electrons is increased, i.e., more scattering channels
  become open.
  Could they tell us
 something about how  the perturbing leads reduce symmetry
   and how an initially
 regular billiard may eventually turn into a chaotic one as
 the number open modes increase?
Below we will argue that nodal points, i.e., the points
 at which the two sets of nodal lines intersect because
 $Re[\psi]=Im[\psi]=0$, carry important information in this respect. Thus we
 will study their spatial distributions and try to characterize chaos in terms
 of such distributions. The question we wish to ask is simply  if
 one can find  a distinct
 difference between  the distributions for nominally regular and
 irregular cavities.

In addition, which other signatures of quantum chaos may one find in
 the coherent transport in  open billiards?
The spatial distribution of nodal points play a decisive role in how
the flow pattern is shaped.
Therefore we will also  study  the general behavior of streamlines derived from
 the probability current associated with a stationary
scattering state
$$
\psi=\sqrt{\rho}\exp{(iS/\hbar)}
$$
The time independent Schr\"{o}dinger equation can be decomposed
as\cite{hirsch1,holland}
$$
E=\frac{1}{2}mv^2+V+V_{QM},
$$
$$
\nabla \rho {\bf v} = 0,\quad m\dot{\bf X}=\nabla S.
$$
The separate quantum streamlines are sometimes referred to as Bohm
trajectories\cite{holland}.
In this
alternative interpretation of quantum mechanics it is thought
 that an electron is a "real" particle that follows a
 continuous and causally defined
trajectory (streamline) with a well defined position ${\bf X}$ with the velocity of the
particle given by the expressions above.

 These equations imply
that the electron moves under the action of a force which is not
obtained entirely
 from the classical potential $V$, but also contains a
"quantum mechanical" potential
$$
V_{QM}=-\frac{\hbar^2}{2m}\frac{\nabla^2 \rho}{\rho}.
$$

This quantum potential is negatively large where the wave function is small, and becomes
infinite at the nodal points of the wave function where $\rho(x,y)=0$. Therefore,
the close vicinity of a nodal point constitutes  a forbidden area for
 quantum streamlines contributing to the net transport from source to drain.
When $\rho$ does not vanish, $ S$ is single valued  and continuous.
However
at the nodal point where $\psi=0$, neither $S$ nor $\nabla S$ is well defined.
The behavior of $S$ around these nodal points is discussed in a
\cite{hirsch1,wu,exner}.
For our study the main important property
of the nodal points of $\psi$ is that  the probability current  flows
described by 'open' streamlines cannot encircle a nodal point. On the contrary,
they are effectively repelled from the close vicinity of the nodal points, in a way
as if these were impurities.

The scattering wave functions $\psi$ are found by solving the Schr\"odinger equation
in a tight-binding approximation with the Neumann boundary conditions outside the
billiards, on a distance over which evanescent modes effectively decay to zero.
The energy of the incident electron is $\epsilon=20$  where
 $\epsilon= 2E_Fd^2m^*/\hbar$ in which $E_F$ is the Fermi energy, $d$ the width
 of the channel, and $m^*$ the effective mass.

\section{Distributions of nodal points}

An inspection of the two sets of nodal lines associated with the real
and imaginary parts of the scattering wave function reveals  the
typical pattern of undirectional, self-avoiding nodal lines found
already by  McDonald and Kauffmann \cite{McDonald} for an isolated, irregular
billiard. However, in our case of a
      complex scattering function the nodal lines
   are not uniquely defined because a multiplication of the wave function by an
    arbitrary constant
  phase factor exp$(i\alpha)$ would yield a different pattern.
   The nodal points, on the other hand, appear to helpful in
   this respect.
 They represent a new aspect of the open system and  will
    obviously remain fixed upon a change of the phase of wave function.
    Here we conjecture that
     the nodal points may serve as
    unique markers which should useful for a quantitative characterization of
    scattering wave functions for open systems.

    To be more specific,  we have considered a large number
    of realizations ('samples') of nodal points associated with
    different kinds of billiards
    and present averaged normalized distributions of nearest distances between
     the nodal points.
    Fig.~\ref{fig1} shows the distributions
  for open Sinai (a), Bunimovich (b) and rectangular billiards (c, d). The
   distributions are obtained
 as an average over 101 different values of energy belonging to a specific
 energy window
in which the conductance undergoes a few oscillations as shown by the insets in
Fig.~\ref{fig1}. The cases (a, b, c) present two channel transmission through
the billiards while the case (d) refers to five channel transmission.
 The rectangular billiard is nominally maximal
 in area with numerical size $210\times 100$ and with width of leads equal to 10.

    It is noteworthy that the distribution of nearest neighbors is distinctly
  different from the corresponding distribution
  for random points in the two-dimensional plane \cite{efros,eggert}
  $$
  g(r)=2\pi \rho r\exp(-\pi \rho r^2),\eqno(1)
 $$
 where a density $\rho$ of random points is related to mean separation
$<r>$ as $\rho=1/4<r>^2$.
This distribution is  shown in  Fig.~\ref{fig1} (a) by the thin line
   indicating an underlying correlation between the nodal points of
   transport wave function through the Sinai Billiard.
    In this sense quantum chaos is not randomness.

      With slight deviations the Bunimovich billiard gives rise to the same
   distributions as the Sinai as shown by Fig.~\ref{fig1} (a,b).
   Analysis of the distributions for lower energies  ($\epsilon\approx 20$,
   one channel transmission)
   gives quite similar universal forms as shown in Fig.~\ref{fig1} (a, b), but
   with more pronounced fluctuations because the number of
   nodal points is less at lower energies. Moreover the average over wider
   energy domains with a finer grid or for higher energies
   gives no visible deviations from
   the distributions in Fig.~\ref{fig1} (a, b).

We considered also the Berry's wave function of a chaotic billiard
    which is accepted as standart measure of quantum chaos \cite{berry}:
      $$
    \psi(x,y)=\sum_j |a_j|\exp[ik(\cos\theta_j x+\sin\theta_j y)+\phi_j]\eqno(2)
    $$
where $\theta_j, |a_j|$ and $\phi_j$ are independent random variables.
    We found that distribution of nearest distances between
     the nodal points of (2) has completely the same form as for
     the Sinai billiard Fig.~\ref{fig1} (a). On the other hand an analysis of
     nodal points of wave function
    $$
    \psi(x,y)=\sum_{k_x,k_y}\exp(ik_x x+k_y y)\eqno(3)
    $$
with $k_x, k_y$ distributed randomly leads to the distribution (1) of
random points.

   To supplement the averaging over energy we have also considered the
   positions of leads. Fig.~\ref{fig2} (a) shows the normalized distribution
   of the nearest distances between nodal points for the Sinai billiard
   obtained  as an average over 101 positions of the input lead.
   As seen this distribution has the same form as  the energy averaged
   Sinai billiard  in Fig.~\ref{fig1} (a).
   In the same way Fig.~\ref{fig2} (b) shows the corresponding case of the Bunimovich
   billiard with an asymmetric input lead to be compared with
   Fig.~\ref{fig1} (b). The unsymmetric arrangement of leads
   allows a larger number of eigenstates of the Bunimovich  to participate in
   the electron transport because symmetry restrictions are
   relaxed\cite{IgorZ}.

   On the basis of Figs.~\ref{fig1} and ~\ref{fig2} and comparison with
   the Berry's wave function (2) we therefore argue
  that there is a universal distribution that characterizes open chaotic
   billiards.  At this stage we conclude  that the
 form of the distributions is not sensitive to the averaging
 procedure, to the number of channels of electron transmission and to the type
 of attachment of leads.  The mathematical form
of the universal distribution constitues an interesting problem that remains
to be solved. So does a derivation of the random distribution associated with wave function
 in eq. (3).

 Let us now turn to the case of the nominally regular rectangular billiard.
 In Fig.~\ref{fig1} (c) the distribution functions are given for the case
 of two-channel transmission with the same energy averaging procedure as  for
 the chaotic billiards.  The nearest neighbor distribution clearly displays
 a peak corresponding to a regular set of nodal points
in contrast to other billiards discussed above. This feature is found even
for very high energies around 250 (five-channel transmission).
Therefore the rectangular dot with the two symmetrically
 attached leads displays considerable stability with respect to regular
 nodal points in contrast to  the chaotic  Sinai and Bunimovich billiards.

As indicated, symmetric leads impose restrictions on how states inside the billiard
are selected and mixed on injection of a particle.
In Fig.~\ref{fig2} (c) the result of averaging over the positions of the input
lead is therefore shown for the
rectangular billiard at a fixed energy chosen from the energy domain in
 Fig.~\ref{fig1} (c). As may be expected  the pronounced peak in the distribution
function of nearest nodal points has now disappeared. Moreover,
the distribution is  close to the case of the Bunimovich billiard
in Fig.~\ref{fig1} (b) and  Fig.~\ref{fig2} (b). Evidently the
non symmetrical positioning of leads disturb the nominally regular billiard
in a much more profound way, effectively rendering it chaotic characteristics.
To reconfirm
this conclusion we have also performed calculations of distribution
 of nodal points within
the same energy domain and the same number of energy steps as in
 Fig.~\ref{fig1} (c) but for non symmetrical positions of the input lead.
In fact, the distribution function of nearest distances in Fig.~\ref{fig2} (d)
demonstrates the close similarity with the position average of the nodal points.
Therefore the non universal behavior of the distribution function of nodal points
for the rectangular billiard shown in Fig.~\ref{fig1} (c, d) is the result of
only a few symmetrical eigenstates taking part in the transmission because of symmetry
restrictions.

In order to give a quantitative  measure of disorder of nodal point
patterns  we consider the Shannon entropy $S$\cite{guarino} normalized for
each specific billiard by the entropy of fully random points.
Numerical values for $S$  are specified in
Figs.~\ref{fig1} and ~\ref{fig2}. As may be expected there is a clear tendency
towards maximal
 entropy
for chaotic billiards for the same energy window.
A similar tendency is clearly seen for the position average  (Fig.~\ref{fig2}).
A case of  rectangular billiard
with entropy 0.95 Fig. ~\ref{fig1} (d) is beyond of this rule
because for the five-channel transmission
 the number of nodal points substantially exceeds other considered
cases irrespective of type of billiard.  Thus the Shannon entropy of
nodal points is important additional
quantitative measure of quantum chaos for the quantum transport through billiards.

\section{Streamlines}

 As mentioned above streamlines are strongly affected by the positions of
  nodal points. Superficially they play the role of impurities.
  It is therfore interesting issue if streamlines
  behave differently for regular and irregular situations and for this reason we
  will consider a few typical examples starting with two well defined systems,
  the nominally regular rectangle and the irregular Sinai billiard.
 Fig. ~\ref{fig3} (a) shows the flow lines in the
 case of the rectangular billiard. The features of the flow lines
 connecting input and output leads are remarkable. It is
clearly seen how the flow  (trajectories) effectively 'channel'
through 'a nodal crystal'  avoiding the individual nodal points.
This picture is evidently very different from semi-classical physics
 and periodic orbit theory\cite{Brack}.
In Fig. ~\ref{fig3} only contributions to the net current are displayed.
In addition there are also vortical motions centered around each nodal point.

The other extreme, the completely chaotic Sinai billiard, is shown in
Fig. ~\ref{fig3} (b). Because  the nodal distribution
is now irregular also the
streamlines
form an irregular pattern when finding their way through the rough potential
landscape. Since a streamline cannot cross itself Fig. ~\ref{fig3} brings to
mind the classical example of meandering rivers in a flat delta landscape.
As well known, slight changes in the topography, for example, by moving only
  a  few obstacles  to new positions, may induce completely new
  flow patterns in a sometimes dramatic ways. In the same way slight variations of the
energy, for example, may affect the quantum streamlines in the Sinai billiard
in an endless way, occasionally forming more collected bunches
connecting the two leads in a more focused way than in Fig. ~\ref{fig3} (b).
The same type of behavior has also been
obtained for a two-dimensional ring in which a tiny variation of external
magnetic flux induce drastic
changes of the flowlines and, as a consequence,
 Aharonov-Bohm oscillations become irregular\cite{pichugin}.

\acknowledgements

This work has been partially supported by the INTAS-RFBR Grant 95-IN-RU-657,
RFFI Grant 97-02-16305 and the Swedish Natural Science Research Council. The
computations were in part performed at the National
Supercomputer Centre at Link\"oping University.


Figure captions\\ \ \\

\begin{figure}
\caption{Normalized distributions for nearest separations between nodal points
 (in units of mean separation) averaged over an energy window
 for the chaotic Sinai (a) and Bunimovich billiards
 (b) and for two rectangular billiards (c, d).
 The Shannon entropy S is given for each separate case. Cases
(a), (b) and (c) correspond to two channel transmission and (d) to
five open channels. The corresponding conductance (in units of $2e^2/h$)  versus energy
are shown in the  insets which also define the energy window for each case.
The distribution (1) for the nearest distances among completely
random points  is shown by thin line in (a).}
\label{fig1}
\end{figure}

\begin{figure}
\caption{Normalized distributions averaged over position of input lead for
the Sinai billiard (a), over an energy window from $\epsilon= 49$ to $50$
 for the Bunimovich billiard
with non-symmetric input lead (b),
over lead positions for the rectangular billiard  (c), and over an energy
window
for the rectangular billiard  with non-symmetric input lead (d).}
\label{fig2}
\end{figure}

\begin{figure}
\caption{Streamlines and  positions of vortices (nodal points) at maximum conductance
$(2e^2/h)$
 for (a) the rectangle with $\epsilon = 20.44$ and (b) for the
Sinai billiard with $\epsilon=20.79$.}
\label{fig3}
\end{figure}

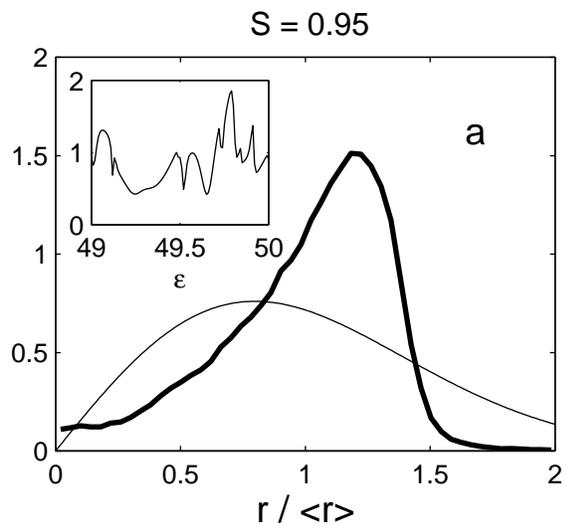
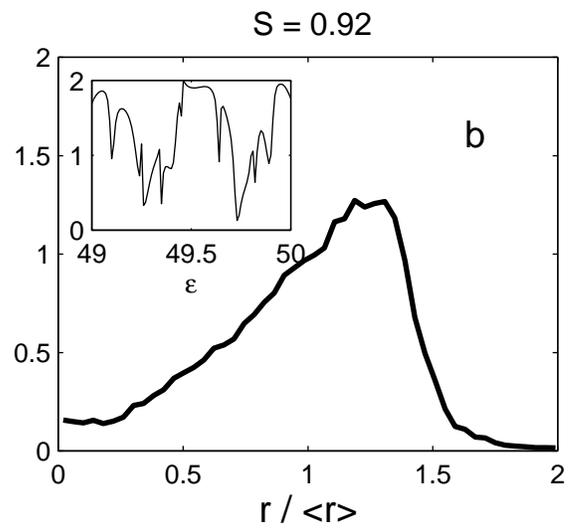
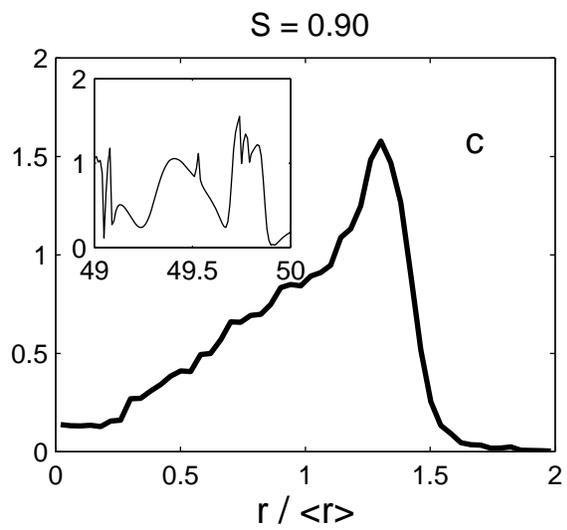
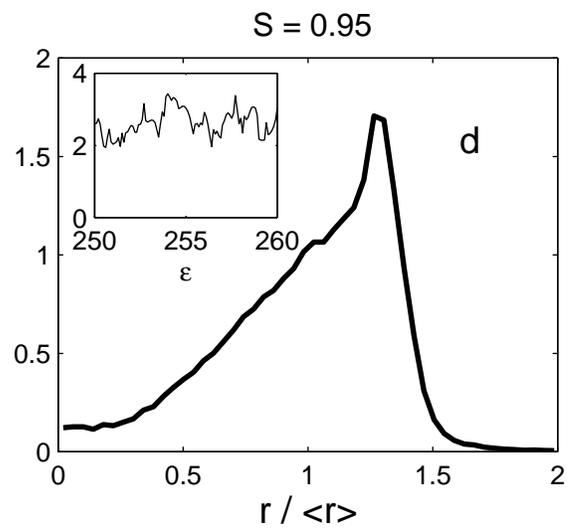
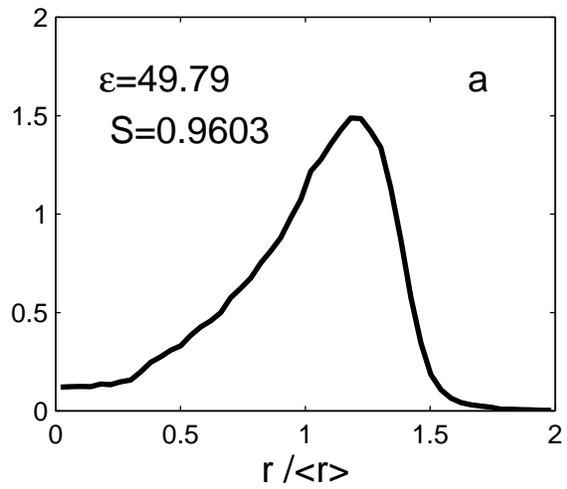
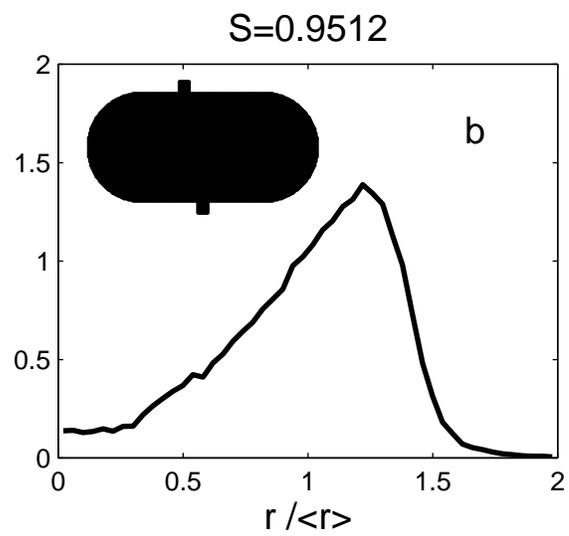
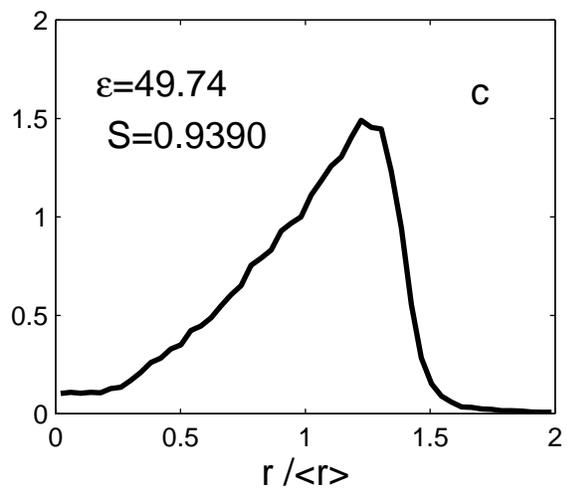
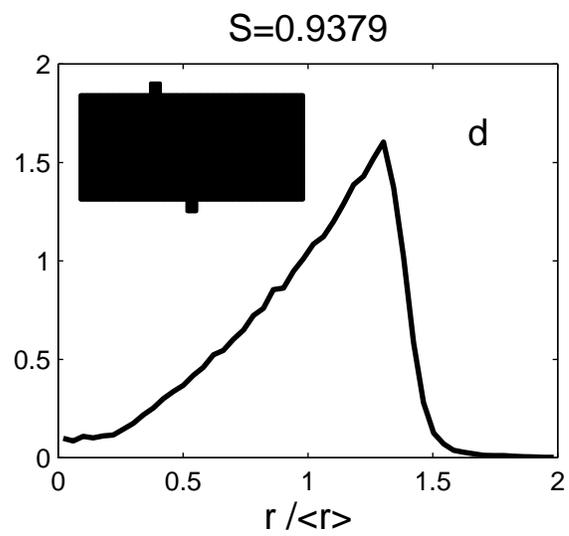
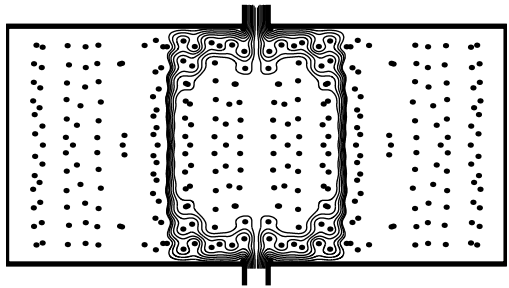
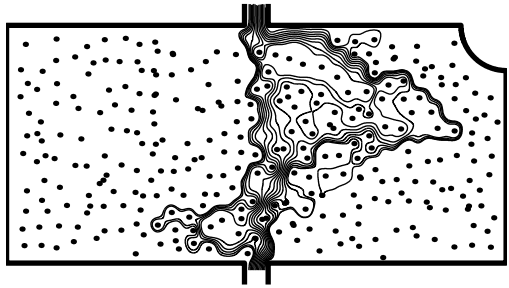
\end{document}